\documentclass{article}

\usepackage{amssymb}

\usepackage{graphicx}

\begin{document}

\title{On the Everett programme and the Born rule.}
\author{Patrick Van Esch \\
\textit{Institut Laue Langevin, Grenoble} \\
 France}
 \label{firstpage}

\maketitle

\begin{abstract}
Proponents of the Everett interpretation of Quantum Theory have
made efforts to show that to an observer in a branch, everything
happens as if the projection postulate were true without
postulating it.  In this paper, we will indicate that it is only
possible to deduce this rule if one introduces another postulate
that is logically equivalent to introducing the projection
postulate as an extra assumption. We do this by examining the
consequences of changing the projection postulate into an
alternative one, while keeping the unitary part of quantum theory,
and indicate that this is a consistent (although strange) physical
theory.

\end{abstract}

\section{Introduction}

An important part in the programme of 'the Everett interpretation'
(first proposed in Everett 1957) or the relative state
interpretation of quantum theory is to avoid postulating the
Projection Postulate (PP) which von Neuman (1955) entitled
'process (1)', so that we can assume that only unitary evolution
is necessary, and that this induces, for an observer, a
measurement history \emph{as if} the Projection Postulate were
true (Dewitt \& Graham, 1973). The relative state view on quantum
theory needs to address two issues in order to produce the effects
of the Born rule: (i) it needs to show \emph{in what basis} an
effective Born rule will emerge and (ii) it needs to show that the
correct probabilities will emerge for an observer. A good recent
overview of the history of the subject can for example be found in
the work of Rubin (2003).

Much progress has been made on (i), mainly through the decoherence
programme, thoroughly described in a book by Joos et al. (2005).
Its relevance for the Everett programme is for instance described
in work by Zurek (1998). However, (ii) seems to be much more
problematic.

As an example of recent work on (ii), Deutsch (1999) has proven an
interesting theorem, which states, under additional 'reasonable'
assumptions, that the only way a rational decider can assign
probabilities to outcomes of future quantum measurements, is
through the Born rule. Several papers then argued on the validity
of this proof (Hanson 2003 ; Wallace 2002, 2003 ; Gill 2003 ;
Greaves 2004).  Finkelstein (2000) claims that Gleason's theorem
solves the issue ; but this theorem also contains an additional
assumption.

We will argue in this paper that the extra assumptions, in all
these cases, are logically equivalent to introducing the PP. This
doesn't affect the value of Deutsch's and other's work, which
allow us to reformulate the PP in other terms, and thus to
understand better what exactly are its essential ingredients. But
it means that there is no hope of deriving the PP directly from
the rest of the machinery of quantum theory, and hence puts that
part of the Everett programme to an end.

The situation is in certain ways reminiscent of attempts, during
more than 2 millennia, of deducing Euclid's Fifth Postulate
(Trudeau, 1987 gives a marvelous account on that history) from the
other postulates of Euclidean geometry, until it was resolved by
Gauss and Bolyai and independently by Lobachebsky, by showing that
there was a consistent way of building Non-Euclidean geometry by
explicitly introducing an alternative Fifth Postulate. We will try
to apply the same strategy: we will postulate an Alternative
Projection Postulate (APP) and see that this gives rise to a
consistent theory on the same level as the standard theory ---
even if the theory is experimentally of course completely wrong.
The very logical existence of this theory then indicates the
independence of the PP from the unitary part of quantum theory.

Apart from proving the logical independence of the PP from the
unitary part of quantum theory, constructing such a logical
alternative has another practical advantage: it allows one more
easily to find out where "proofs" of the PP make hidden (or
explicit) extra assumptions. We will then examine where exactly it
is in disagreement with Deutsch's 'reasonable assumptions', or
with Gleason's theorem.

\section{The Alternative Projection Postulate.}

Let us propose a quantum theory \`{a} la von Neumann (1955),
except for the projection postulate (which he calls process (1)),
which we replace by the Alternative Projection Postulate (APP).
It has to be said that the APP can seem slightly more limited in
scope than the original PP, in that only measurements with a
finite number of different outcomes are handled.  However, this is
not a physical shortcoming, because any true measurement can
result only in a finite amount of information, and hence in a
finite number of discrete outcomes. We propose the APP:

\begin{quote}
Let $\{\hat{X_k}\}$ be a set\footnote{We need a set, only because
we want to be able to label the outcomes with several different
real numbers.  As long as there are only a finite number of
different outcomes, one single operator could in principle be
sufficient.} of commuting self-adjoint operators with a finite,
common discrete spectrum (the different possible outcomes of the
measurement).  This finite spectrum is given by a finite series of
sets of eigenvalues $\{x^k\}_i$.  The full set of $\{\hat{X_k}\}$
defines the measurement to be performed.  To each different set of
eigenvalues $\{x^k\}_i$ of $\{\hat{X_k}\}$ corresponds a projector
$P_i$ on the space of common eigenvectors belonging to
$\{x^k\}_i$. There are by hypothesis only a finite number of such
projectors, which form a complete, orthogonal set \footnote{This
means that $\sum_i P_i = 1$ and that $P_i P_j = 0$ if $i \neq
j$.}. Let $N$ be the (finite) number of projectors $P_i$.
 Let $|\psi\rangle$ be the state of the system in
Hilbert space before the measurement. Let $n_{\psi}$ be the number
of projectors for which $P_i|\psi\rangle$ is different from 0. We
obviously have: $1 \le n_{\psi} \le N$.  If the system is in state
$|\psi\rangle$, each of the set of values $\{x^k\}_i$
corresponding to such a projector has a probability $1/n_{\psi}$
to be realized. The other sets of values have probability 0 to be
realized. If the outcome of the measurement equals $\{x^k\}_u$,
then the state after measurement equals $P_u |\psi\rangle$,
properly normalized.
\end{quote}

One notices the difference with the original PP as found in most
standard texts on quantum mechanics, such as Cohen-Tannoudji
(1997): the probability equals $1/n_{\psi}$ instead of
$\langle\psi|P_i|\psi\rangle$\footnote{We could even go further
and postulate the probability to be equal to $\alpha/n_{\psi} +
(1-\alpha)\langle\psi|P_i|\psi\rangle$ with $\alpha$ a real number
between 0 and 1, which defines a generalized APP for each value of
$\alpha$.}.

We should point out that the APP is in fact the most natural
probability rule that goes with the Everett interpretation: on
each "branching" of an observer due to a measurement, all of its
alternative 'worlds' receive an equal probability.

\section{Consistency of Quantum Theory based upon the APP.}

The alternative quantum theory (which is normal quantum theory,
with the PP replaced by the APP, for short AQT) will turn out to
be a physical theory which is completely different from standard
quantum theory (SQT) and also experimentally totally wrong.
However, we will try to show that it is a consistent theory on the
same level as SQT.  It is a priori very difficult to prove that a
physical theory is consistent.  However, the bulk of the
mathematical machinery of SQT and AQT is the same (the unitary
evolution).  The intervention of the APP on the mathematical
machinery is the same as the PP (indeed, it is a projection of the
state vector on an eigenspace, followed by a normalization of the
projection, in both cases).  So on the purely mathematical side,
both theories are identical concerning the evolution of the state
vector.

The subtler aspects are related the physical interpretation.
Indeed, the PP is the only link to experimental quantities, and
this is replaced by the APP.  We have to ensure that through the
APP, we arrive at an operational definition of the mathematical
entities which is consistent.  We will show that it is in fact
exactly the same as in SQT. Furthermore, we have to prove that
different mathematical descriptions describing the same physical
situation give identical results.  This means invariance under
unitary transformations, and invariance under different ways of
formulating the same measurement process.

\subsection{Respect of unitary transformations.}

The representation of the state space, and all of the unitary
evolution machinery, can undergo a unitary transformation without
changing their interpretation.  We have to ensure that our AQT
gives identical results when such an isomorphism is applied.  So
we need to show:

\begin{quote}
Any unitary transformation of the Hilbert space of states, such
that $|\psi\rangle$ is mapped upon $U|\psi\rangle$ and every
observable $O$ is mapped upon $UOU^{\dagger}$, leaves the results
and effects of measurements, such as they are introduced by the
APP, invariant.\end{quote}

The proof is straightforward. First of all, the projectors $P_i$
are transformed into $UP_iU^{\dagger}$, so that the projections of
$U|\psi\rangle$ are transformed into $U P_i |\psi\rangle$. This
projection is zero if and only if $P_i|\psi\rangle = 0$, so the
number of non-zero projections $n$ is conserved, as well as the
eigenvalues $\{x^k\}_i$ which belong to such projections.  The
probabilities of the measurement results are hence the same before
and after the transformation. Also any further evolution, after
the measurement, is equivalent to the evolution before
transformation, given that the state after the measurement (with
result $\{x^k\}_u$) is now $U P_u |\psi\rangle$, properly
normalized, which is nothing else but the transformation, under
$U$, of the state we would have obtained under the same
circumstances.

\subsection{Measurement results predicted with
certainty in AQT and SQT are the same.}

In SQT, the interpretation of the mathematical entities (state
vector, observable etc...) is completely fixed by the experimental
results predicted with certainty.  We will show that this
interpretation is exactly the same in AQT.

\begin{quote}
If $|\psi\rangle$ is an eigenvector of the different $\hat{X}^k$
with respective eigenvalues $\{x^k\}_i$, then the measurement will
give with certainty the result $\{x^k\}_i$ for this observable,
and the state after measurement will still be
$|\psi\rangle$.\end{quote}

The proof is trivial and based upon the fact that $P_i
|\psi\rangle = |\psi\rangle$ and $P_j |\psi\rangle = 0$ for $i
\neq j$.

We also have:

\begin{quote}
If $|\psi\rangle$ doesn't have any component with eigenvalues
$\{x^k\}_i$, then the measurement will give a result which is
different from $\{x^k\}_i$, with certainty.\end{quote}

This is a direct consequence of the APP.

\subsection{Equivalence of physically identical measurements.}

Two measurements are physically identical if exactly the same
information is extracted by either of both measurements.  We will
show that two mathematically different sets of observables,
$\{X_k\}$ and $\{Y_k\}$, which correspond to physically identical
measurements, result in operationally identical results.

If $\{X_k\}$ and $\{Y_k\}$ extract the same information, this
means that to each distinct set of eigenvalues $\{x^k\}_i$
corresponds exactly one set of distinct eigenvalues $\{y^k\}_i$
and vice versa ; and that for each case where the result of
measurement of $\{X_k\}$ gives with certainty $\{x^k\}_i$, then
the result of the measurement if we measure $\{Y_k\}$ should give
with certainty $\{y^k\}_i$.  But this means that $P_i^X = P_i^Y$.
As only the projectors play a role in the APP, the two
measurements with equivalent sets of observables yield exactly the
same results under the APP.

\section{Strange properties of AQT.}

We will discuss some strange properties of AQT, which immediately
disqualify it as a possible candidate of a physical theory of our
world. However, we want to emphasize that such a world is a
logical possibility including the unitary part of quantum theory,
even if it is a very strange one to our standards. By examining
some very strange and 'unreasonable' properties, we also show how
easy it is to eliminate AQT by introducing 'reasonable
assumptions'.

If we take the topology induced by the Hilbert in product, then an
arbitrary small change of $|\psi\rangle$ (by adding a small
component of an eigenspace that was orthogonal to $|\psi\rangle$)
can induce a discrete change in probabilities of outcomes. But
this, as such, is not an internal contradiction of the theory.
Note also that in general, the state of a system is never strictly
orthogonal to an eigenspace of a set of observables (except
immediately after measurement), so one can usually assume that $N
= n$, except immediately after a measurement.

This also means that if there is a small time lapse between two
identical measurements, that the two results are uncorrelated
except in the case where all the $\{X^k\}$ commute with the full
Hamiltonian of the system.  Although this result seems very
strange indeed, it does not necessarily indicate an internal
inconsistency, but just means that measurements incompatible with
the full $H$ are a waste of time, because the information is
immediately lost.  As we usually don't know the full $H$, this
means that most measurements are a waste of time. AQT describes a
very random world indeed, in which, most of the time, the outcomes
of measurements are independent of the state the system is in!

Another strange property of AQT is the following.  Imagine that we
consider two different, commuting observables, measuring the same
quantity. We have a course-grained one, $X$, and a fine-grained
one, $Y$. Let us assume that $Y$ has 5 distinct eigenvalues,
namely 1,2,3,4, and 5. Let us assume that $X$ has two eigenvalues,
10 and 20, and that $X$ takes on value 10 when $Y$ takes on value
1, and that $X$ takes on the value 20 when $Y$ takes on the values
2,3,4, and 5. For a general state $|\psi\rangle$ which isn't
'particular' with respect to $X$ or $Y$ (meaning, has non-zero
components for all of their eigenstates), a measurement consisting
purely of $X$ will result in a probability 0.5 for value 10, and
0.5 for value 20.  A measurement consisting purely of $Y$ will
give a probability of 0.2 for each of (1,2,3,4 and 5), and hence,
if we calculate X from it, a probability of 0.2 for finding 10 and
a probability 0.8 for finding 20.  So, depending on whether we
measure also $Y$ or not, the result $x = 10$ has a probability of
0.5 or 0.2.  At first sight, this is a strange result ; however it
is not an inconsistency, and just extends the "strangeness"
already present in SQT.  In SQT, incompatible measurements
influence each other's outcome probabilities ; in AQT, even
compatible, but different, measurements influence each other's
otucome probabilities. Indeed, the measurement consisting purely
of ${X}$ is not physically equivalent with the measurement
consisting of ${X,Y}$, because a different amount of information
is extracted. On the other hand, the measurement ${X,Y}$ and the
measurement ${Y}$ are identical, because the measurement of $Y$ is
also a measurement of $X$ ; this is an illustration of the
equivalence of measurements. It is a property of AQT that changing
the resolution of a measurement can change the probabilities of
the outcome of the crude measurement, which is not the case under
SQT.  Note that this property of SQT is the 'non-contextuality'
needed in the application of Gleason's theorem.  Indeed, in AQT,
the fact that the probability of measuring 10 for $X$ depends on
whether we have measured $Y$ or not (and which corresponds
physically to two different measurement situations), means that
AQT is not non-contextual.  So an axiom of non-contextuality can
be seen as an axiom, equivalent to the Born rule.

We now see where some of Deutsch's 'reasonable' assumptions (made
more explicit in Wallace (2003)) explicitly rule out AQT. One
instance is requiring identical probabilities under Payoff
Equivalence, when the function $f(x_i)$ is not invertible
(meaning, for $x_i \neq x_j$ we can have $f(x_i) = f(x_j)$.
Indeed, in AQT, the measurement $f(X)$ and $X$ are not considered
equivalent because $f(X)$ extracts less information from the
system than $X$. Using payoff equivalence with $f$ non-bijective
is a crucial point in the proof of Deutsch's theorem, as made
clear in steps (35) and (36) in Wallace (2003).

We want again to emphasize that these strange results are a
logical possibility of a theory evolving according to "unitary
quantum theory".  They are simply different from those given by
SQT, in the same way that geometrical results in hyperbolic
geometry are different from the geometrical results by Euclidean
geometry, and are "strange" as compared to everyday "geometrical
measurements".

\section{Discussion}

In this paper we tried to show that quantum theory, with the
projection postulate replaced by an alternative one, gives rise to
a consistent physical theory, at least at the same level as
standard quantum theory. This theory has very strange consequences
and can certainly not describe our world, but its consistency (in
relation to standard theory) proves that it is not possible to
deduce the projection postulate from the 'unitary' part of quantum
theory, in the same way it is not possible to deduce Euclid's
Fifth Axiom from the four other ones. All attempts to do so by
introducing extra assumptions just indicate that those extra
assumptions are logically equivalent to the projection postulate.
This, by itself, is not necessarily a meaningless exercise.

\label{lastpage}

\end{document}